\documentclass[12pt,preprint]{aastex}

\usepackage{amsmath,amssymb}

\shorttitle{Compton scattering in the Klein-Nishina regime}

\shortauthors{Kusunose, Takahara}

\begin{document}

\title{Compton scattering in the Klein-Nishina Regime Revisited}

\author{Masaaki Kusunose}
\affil{Department of Physics, School of Science and Technology,
Kwansei Gakuin University, Sanda 669-1337, Japan}
\email{kusunose@kwansei.ac.jp}

\and

\author{Fumio Takahara}
\affil{Department of Earth and Space Science, 
Graduate School of Science, Osaka University,
Toyonaka 560-0043, Japan}
\email{takahara@vega.ess.esi.osaka-u.ac.jp}

\begin{abstract}
In blazars such as 3C 279, GeV gamma-rays are thought to be produced
by inverse Compton scattering of soft photons injected from external
sources into the jet.  Because of the large bulk Lorentz factor of the jet, 
the energy of soft photons is Doppler shifted in the comoving
frame of the jet, and the scattering is likely to occur 
in the Klein-Nishina regime.
Although the Klein-Nishina effects are well known, 
the properties of the electron and emission spectra have not
been studied in detail in the environment of blazars.
We solve the kinetic equation of electrons with the spatial escape term
of the electrons to obtain the electron energy spectrum in the jet 
and calculated the observed emission spectrum.
In calculations of the Compton losses in the Klein-Nishina regime,
we use the discrete loss formalism to take into account the significant
energy loss in a single scattering.
Although the scattering cross section decreases because of
the Klein-Nishina effects, ample gamma rays are emitted by
inverse Compton scattering.
When the injection spectrum of electrons obeys a power law, 
the electron spectrum does not follow a broken power law, 
as a result of the Klein-Nishina effects,
and a large number of high-energy electrons remain in the emitting region.
\end{abstract}


\section{Introduction}

Inverse Compton (IC) scattering is an important process 
in producing high-energy photons
in astrophysical objects such as active galactic nuclei and gamma-ray bursts.
Blazars are one of the important sites at which
IC scattering is a major radiation process.
The radiation from blazars is well described by
synchrotron emission and IC scattering \citep[e.g.,][]{sbr94,it96,gmd96}.
The spectral energy distribution of blazars has two peaks in the 
$\nu$-$\nu F_\nu$ representation; 
one is in the range between IR and X-rays, and the other, 
in gamma rays such as GeV and TeV regions.
The emission region of blazars is thought to be in the relativistic jet close
to the central black hole, and
the nonthermal electrons accelerated in the jet are probably
responsible for the emission.
Synchrotron radiation emitted by the nonthermal electrons produces
the low-energy peak, and IC scattering by the same nonthermal electrons 
produces the high-energy peak.
Although there are hadronic models that assume relativistic protons
\citep[e.g.,][]{muc03}, we confine ourselves to leptonic models,
which assume radiation from nonthermal electrons/positrons alone.

In the jet of 3C 279, nonthermal electrons Compton scatter 
external soft photons to produce GeV gamma rays \citep{it96}; 
the source of the external soft photons
may be accretion disks or broad-line regions \citep[e.g.,][]{ds93,sbr94}.
In a previous paper \citep{kus03}, we showed that the electron
spectrum in the emission region of 3C 279 does not obey a conventional broken 
power law, which appears if scattering occurs in the Thomson regime.
The electron energy spectrum of 3C 279 is found to be
harder than the broken power law in the high-energy region
because of the Klein-Nishina (KN) effects.

When the electron Lorentz factor $\gamma$ and the seed soft photon energy
normalized by the electron rest mass $x = h \nu / (m_e c^2)$
satisfy $\gamma x < 1$, IC scattering occurs in the Thomson regime.
Here $m_e$ and $c$ are the electron rest mass and the speed of light,
respectively.
On the other hand, if $\gamma x > 1$, 
IC scattering occurs in the KN regime \citep[e.g.,][]{bg70,rl79}.
In astrophysical objects high-energy radiation is often emitted by
nonthermal electrons with a power-law distribution.
In the objects in which nonthermal electrons are injected, radiate, and escape,
the electron spectrum becomes steeper for $\gamma > \gamma_\mathrm{br}$,
where $\gamma_\mathrm{br}$ is the Lorentz factor determined
by the balance between the Compton cooling and the electron escape, 
if the scattering occurs in the Thomson regime.
When the injection spectrum obeys a power law,
$q_e(\gamma) \propto \gamma^{-p}$, 
the steady state spectrum obeys a broken power law,
$n_e(\gamma) \propto \gamma^{-p}$ for $\gamma < \gamma_\mathrm{br}$
and $n_e(\gamma) \propto \gamma^{-p-1}$ for $\gamma > \gamma_\mathrm{br}$.

The effects of IC scattering
in the KN regime have been studied by several authors.
\cite{blu71}, for example, showed that the electron spectrum becomes
harder in the KN regime, assuming that monoenergetic
soft photons are scattered.
\cite{zk93} demonstrated that when the electrons are injected
with a power-law distribution
with the power-law index of 2, the emission spectrum is flat in
the $\nu$-$\nu F_\nu$ representation if the KN effects are
taken into account, although they assumed that the scattering cross
section is zero for $\gamma x > 3/4$ to simulate the KN effects.
These properties were applied to the explanation of 
the TeV emission from Mrk 421 and its flat spectrum in the GeV$-$TeV region
in the $\nu$-$\nu F_\nu$ representation.
\cite{zdz89} solved kinetic equations for various types of the injection 
spectra of electrons and soft photons, using the KN cross section.

In above work \citep{blu71,zdz88}, the electron escape from the emission 
region was not included.
In objects such as blazars, the emission region exists in the relativistically
moving jets.  
Because the advection or adiabatic expansion of nonthermal particles 
removes the particles from the emission region on a timescale
close to the light crossing time $R/c$, where $R$ is the radius of the
emission region,
the effect of escaping particles on the spectra of electrons
and photons is important.
In this paper we show the properties of the spectra of electrons and photons 
when the electron escape is included and
the scattering occurs dominantly in the KN regime.
We also assume that synchrotron radiation is not important for 
radiative cooling; this is justified when ample soft photons from
external sources are injected as seed photons of IC scattering.

In \S \ref{sec:ke} the kinetic equation of electrons is presented, 
and numerical results are given in \S \ref{sec:nr}.
Finally we summarize our results in \S \ref{sec:sum}.


\section{Kinetic Equation} \label{sec:ke}

The kinetic equation for the steady state electrons
is given by
\begin{equation}
-n_e(\gamma) \int_1^\gamma d \gamma^\prime \, C(\gamma, \gamma^\prime)
+ \int_\gamma^{\infty} d \gamma^\prime n_e(\gamma^\prime) \,
C(\gamma^\prime, \gamma) + q_e(\gamma) 
- \frac{n_e(\gamma)}{t_\mathrm{esc}} = 0 \, ,
\label{eq:kin}
\end{equation}
where $n_e(\gamma)$ is the number density of electrons per unit $\gamma$,
$q_e(\gamma)$ is the injection rate of nonthermal electrons
per unit volume and $\gamma$, $t_\mathrm{esc}$ is the electron escape time,
and $C(\gamma, \gamma^\prime)$ describes the transition rate of 
an electron from $\gamma$ to $\gamma^\prime$.
We assume that $q_e(\gamma)$ is given by
\begin{equation}
q_e(\gamma) = q_0 \, \gamma^{-p} \, \exp (-\gamma/\gamma_\mathrm{max}) \, ,
\quad \gamma > \gamma_\mathrm{min} \, ,
\end{equation}
where $p$, $\gamma_\mathrm{min}$, $\gamma_\mathrm{max}$, 
and $q_0$ are parameters.

In scattering the electron Lorentz factor $\gamma$ decreases 
to $\gamma^\prime$.
When the number spectrum of soft photons per unit volume, 
$n_\mathrm{soft}(x)$, is given,
the rate of scattering ($\gamma \rightarrow \gamma^\prime$) per electron
is given by Jones (1968; see also Zdziarski 1988) as
\begin{equation}
\label{eq:c}
C(\gamma, \gamma^\prime) = \frac{3 \sigma_\mathrm{T} \, c}{4}
\int_{E_*/\gamma}^\infty  \frac{1}{E \gamma} \left[
r + (2-r)\, \frac{E_*}{E} - 2 \left(\frac{E_*}{E}\right)^2
- 2 \, \frac{E_*}{E} \, \ln \frac{E}{E_*} \right] \, n_\mathrm{soft}(x) \, 
d x \, ,
\end{equation}
where
\begin{equation}
E = \gamma \, x \, ,  \quad
E_* = \frac{\gamma/\gamma^\prime - 1}{4} \, , \quad E > E_* \, ,
\quad 
r = \frac{1}{2} \left(\frac{\gamma}{\gamma^\prime} 
+ \frac{\gamma^\prime}{\gamma} \right)  \, .
\end{equation}
Here $\sigma_\mathrm{T}$ is the Thomson cross section, 
and we assume that the soft photons are isotropic.
Equation (\ref{eq:c}) retains the zeroth-order term in a double 
expansion of the exact rate in $1/\gamma^2$ and $x/\gamma$;
the exact cross section is given by Jones (1968; see also Coppi \&
Blandford 1990).
Since we are interested in the KN regime, 
the above approximated cross section 
is enough to evaluate the electron spectrum for $\gamma \gg 1$.
We assume that soft photons with energy $x_{s}$ are isotropically
distributed around the jet, where $x_{s}$ is measured
in the frame of the soft photon source.
In the jet frame, the soft photon energy is Doppler shifted to 
$x = \Gamma x_{s}$, 
where $\Gamma$ is the bulk Lorentz factor of the jet.
The radiation field is approximated to be isotropic in the jet frame
for simplicity.

To solve equation (\ref{eq:kin}), we discretize the equation as
\begin{equation}
-n_e(\gamma_i) \sum_{j=1}^{i-1} C(\gamma_i, \gamma_j) \gamma_j 
\Delta \ln \gamma 
+ \sum_{j=i+1}^{J} n_e(\gamma_j) C(\gamma_j, \gamma_i) \gamma_j 
\Delta \ln \gamma 
+ q_e(\gamma_i) - \frac{n_e(\gamma_i)}{t_\mathrm{esc}} = 0 \, ,
\end{equation}
where $\gamma_i$ ($i = 1, \cdots, J$) is binned logarithmically
and  $\Delta \ln \gamma = \ln \gamma_{i+1} - \ln \gamma_{i}$
with $\Delta \ln \gamma$ being constant 
(typically $\Delta \ln \gamma = 2 \times 10^{-3}$).
Here we omitted the term $C(\gamma_i, \gamma_i)$, because
most electrons lose energy in scattering.
We then obtain
\begin{equation}
\label{eq:discr}
n_e(\gamma_i) = A^{-1} \left[ \sum_{j = i+1}^{J} n_e(\gamma_j) 
C(\gamma_j, \gamma_i) \gamma_j \Delta \ln \gamma
+ q_e(\gamma_i) \right] \, ,
\end{equation}
where
\begin{equation}
A = \sum_{j = 1}^{i-1} C(\gamma_i, \gamma_j) \gamma_j
\Delta \ln \gamma + \frac{1}{t_\mathrm{esc}} \, .
\end{equation}
We calculate $n_e(\gamma_i)$ numerically from equation (\ref{eq:discr})
with the upper boundary condition such that $n_e(\gamma_J) = 0$.
Because $C(\gamma, \gamma^\prime)$ in equation (\ref{eq:c}) is applicable
to large values of $\gamma$, our numerical calculations are limited
to $\gamma_1 >$ several.

The injected electrons lose energy by IC scattering
and move to the lower energy region in the energy space.
The cooling rate of IC scattering is given by
\begin{equation}
\dot{\gamma} = \int_1^\gamma d \gamma^\prime \,
(\gamma^\prime - \gamma) \, C(\gamma, \gamma^\prime) \, .
\end{equation}
The cooling time is then defined as
$t_\mathrm{IC}(\gamma) = \gamma/|\dot{\gamma}|$.
In the Thomson regime,
$t_\mathrm{IC}(\gamma)$ decreases with increasing $\gamma$ and
attains the minimum value at $\gamma \sim x_0^{-1}$,
where $x_0$ is the characteristic energy of soft photons.
For $\gamma > x_0^{-1}$, 
where the scattering occurs in the KN regime,
$t_\mathrm{IC}(\gamma)$ increases.

Nonthermal particles in the jet may escape from the emission region 
by advection or diffusion,
or adiabatic expansion effectively removes them.
Because of the lack of knowledge on the electron escape time,
$t_\mathrm{esc}$ is assumed to be independent of $\gamma$ and given by
$ t_\mathrm{esc} = \zeta_\mathrm{esc} \, \min\{t_\mathrm{IC}(\gamma)\}$,
where $\zeta_\mathrm{esc}$ is a parameter and
$\min\{t_\mathrm{IC}(\gamma)\}$ is the minimum value of
$t_\mathrm{IC}(\gamma)$, which is taken at $\gamma \sim  x_0^{-1}$.

For comparison with the scattering in the KN regime,
we also calculate the electron spectrum derived by 
the Thomson approximation.
The kinetic equation of electrons in the Thomson regime is given by
\begin{equation}
\frac{d}{d \gamma} [ \dot{\gamma}_\mathrm{T}(\gamma) \, n_e(\gamma) ] 
= q_e(\gamma) - \frac{n_e(\gamma)}{t_\mathrm{esc}} \, ,
\label{eq:th-kin}
\end{equation}
where the cooling rate is calculated \citep{lz87}
\begin{equation}
\dot{\gamma}_\mathrm{T} (\gamma) 
= - \sigma_\mathrm{T} c \left( \frac{4}{3} \, \gamma^2 -1 \right)
\int_0^{3/(4 \gamma)} n_\mathrm{soft}(x) \, x \, d x \, .
\end{equation}
Here the scattered photon energy $x^\prime$ is assumed to be given by
$x^\prime = (4/3) \gamma^2 x$, and the scattering cross section is
assumed to be zero for $x \gamma > 3/4$ to simulate the scattering
in the KN regime.
The solution of equation (\ref{eq:th-kin}) is given by
\begin{equation}
n_e(\gamma) = - \frac{1}{\dot{\gamma}_\mathrm{T}(\gamma)} \int_\gamma^\infty
q_e(\xi) \, \exp \left[ \frac{1}{t_\mathrm{esc}}
\int_\gamma^\xi \frac{d \eta}{\dot{\gamma}_\mathrm{T}(\eta)} \right] d\xi \, .
\label{eq:thmsn}
\end{equation}


\section{Numerical Results}
\label{sec:nr}

\begin{figure}
\epsscale{0.8}
\plotone{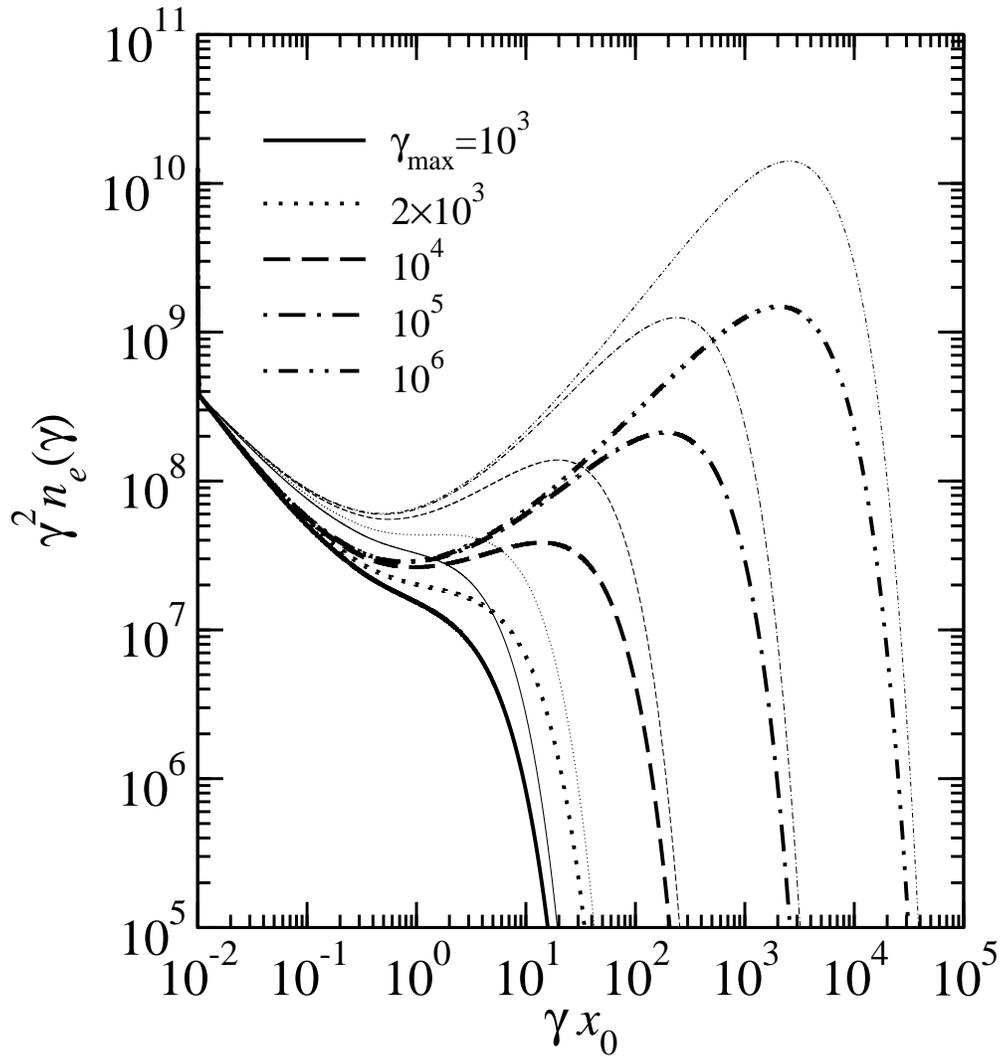}
\caption{Electron spectra for various values of $\gamma_\mathrm{max}$ 
without particle escape from the emission region.
Thick and thin lines are for $p = 2$ and 1.8, respectively.
The injection of monochromatic soft photons with
$m_e c^2 x_{0} = 1.25 \text{keV}$ is assumed.
The KN effects are apparent for $\gamma x_0 \gtrsim 0.1$.}
\label{fig:el-noesc-p2-1.8}
\end{figure}

We solve equation (\ref{eq:kin}) numerically to obtain the electron spectrum
and calculate the emission spectrum.
The emission region is assumed to be in a relativistically moving blob,
and we calculate the emission spectrum according to \cite{gkm01},
who includes the effects of the beaming and the KN cross section,
because our interest is in the application of our model to blazars.
We first solve equation (\ref{eq:kin}) without the term of electron escape
to compare with the results with the escape term.
In Figure \ref{fig:el-noesc-p2-1.8} we show the electron spectra for 
different values of $\gamma_\mathrm{max}$ and $p = 2$ and 1.8
($p = 1.8$ was used in our model of 3C 279; Kusunose et al. 2003).
In this figure, $n_\mathrm{soft}(x) = n_0 \delta( x - x_{0})$
and $x_{0} m_e c^2= 1.25 \, \text{keV}$,
where $\delta$ is the Dirac delta function;
when the bulk Lorentz factor of the jet is $\Gamma = 25$, 
this soft photon energy is 50 eV in the frame of the soft photon source.
The value of $q_0$ is chosen to set the injection rate of electrons to
1 $\text{cm}^{-3} \, \text{s}^{-1}$, and the value of $n_0$ is fixed to
set the energy density of soft photons to $8.3 \times 10^{-2}$ ergs cm$^{-3}$;
this value of the energy density corresponds to 
$1.3 \times 10^{-4}$ erg cm$^{-3}$ 
in the frame of soft photon source if $\Gamma = 25$.
When $p =2$, $n_e(\gamma) \propto \gamma^{-3}$
for $\gamma x_0 \lesssim 0.1$,
i.e., scattering occurs in the Thomson regime.
The KN effects appear for $\gamma x_0 \gtrsim 0.1$,
and the electron spectra become harder.
For example, when $\gamma_\mathrm{max} = 10^6$,
$n_e(\gamma) \propto \gamma^{-1.3}$ for $p =2$ and $10 < \gamma x_0 < 10^3$,
with a prominent peak at $\gamma x_0 \sim 3 \times 10^3$.
It should be noted that if we solve equation (\ref{eq:kin}) at the
lower energy region down to $\gamma = 1$, the electron spectrum should 
have an infinite peak at $\gamma = 1$ because of cooling.
Since we solved equation (\ref{eq:kin}) at large values of $\gamma$,
this peak does not appear in the figures.

\begin{figure}
\plotone{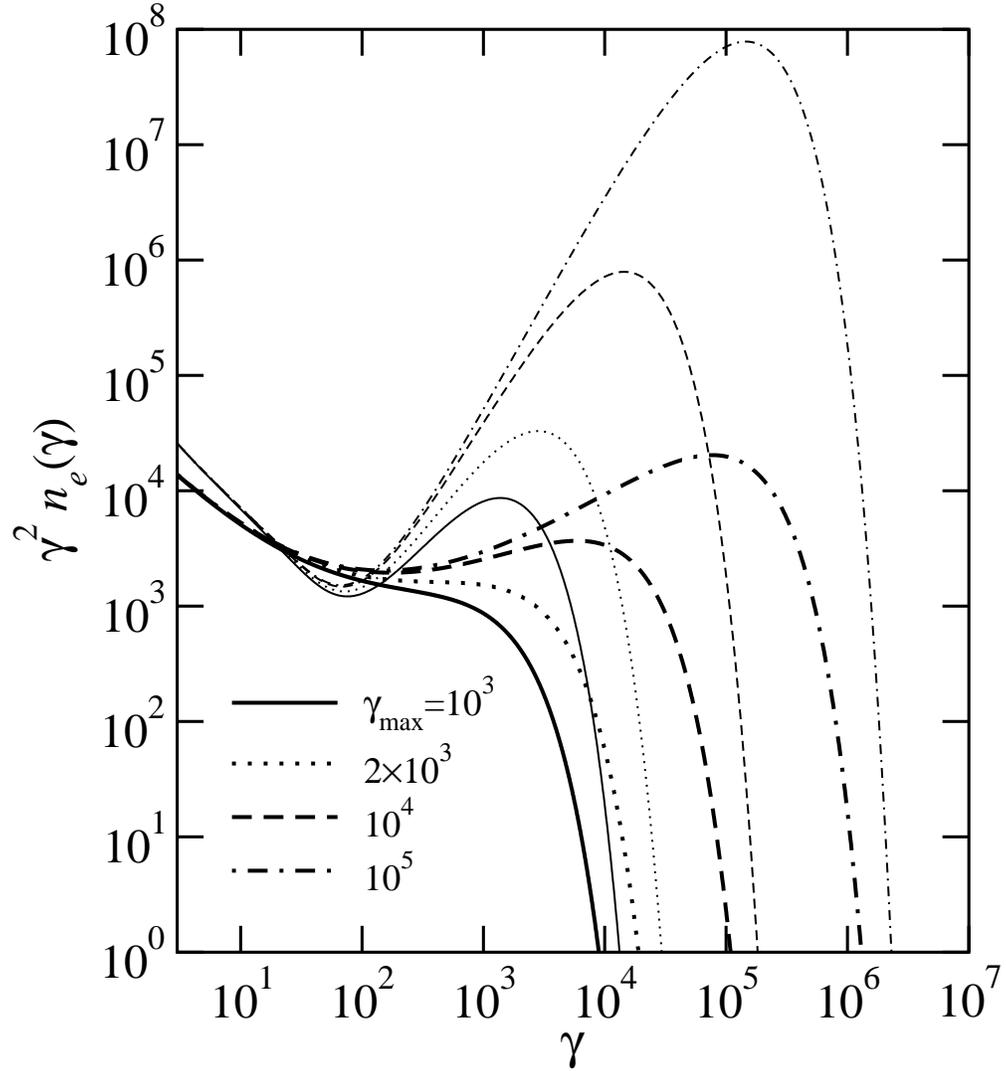}
\caption{Electron spectra calculated with the KN cross section
({\it thick lines}) and with the Thomson approximation ({\it thin lines}).
Injected electrons obey a power law with $p = 2$, and the external
soft photons follow a blackbody distribution 
with a temperature of $1.25$ keV.
}
\label{fig:el-noesc-p2-bb}
\end{figure}

The deviation of the Thomson approximation (eq. [\ref{eq:thmsn}]) from
the KN formulation (eq. [\ref{eq:kin}]) is shown in
Figure \ref{fig:el-noesc-p2-bb}.  
Here the injected soft photons obey the Planck
distribution with a temperature of 1.25 keV and an energy density of
$8.3 \times 10^{-2} \, \text{ergs} \, \text{cm}^{-3}$.
The values of $n_e(\gamma)$ are different from those in 
Figure \ref{fig:el-noesc-p2-1.8}, 
because the external soft photon spectra are different.
Since electrons with $\gamma > 3/(4 x)$ do not suffer from
cooling in the Thomson approximation, 
a large number of electrons stay in the high-energy region.
Thus, just assuming that the cross section vanishes for $\gamma > 3/(4 x)$
considerably overestimates the number density of electrons 
in the high-energy region.
More accurate treatments are needed to properly estimate 
the high-energy electron spectrum.

\begin{figure}
\epsscale{0.8}
\plotone{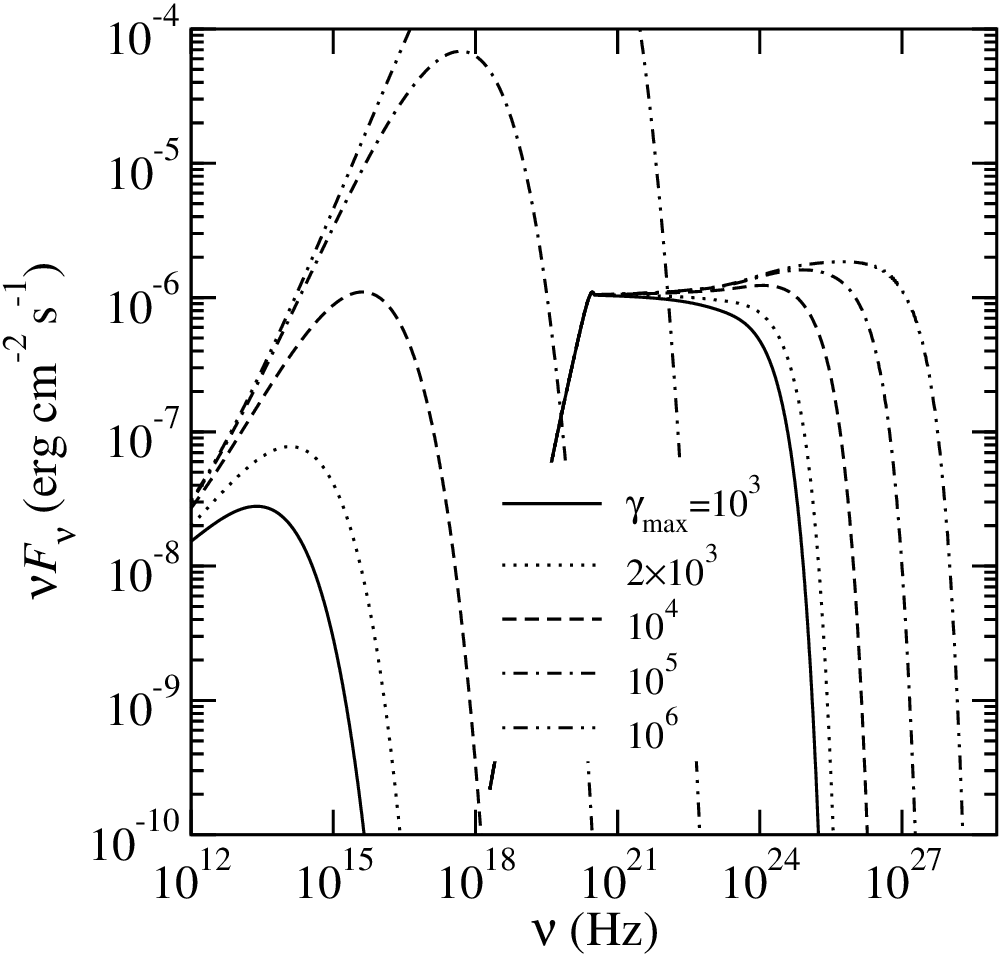}
\caption{Photon spectra emitted by electrons with $p = 2$ 
shown in Fig. \ref{fig:el-noesc-p2-1.8}
and various values of $\gamma_\mathrm{max}$.
The spectra are shown in the
observer's frame.  The curves on the right are the emission by
IC scattering and the curves on the left are by synchrotron radiation
with $B = 0.3$ G.}
\label{fig:ph-noesc-p2}
\end{figure}
\begin{figure}
\epsscale{0.8}
\plotone{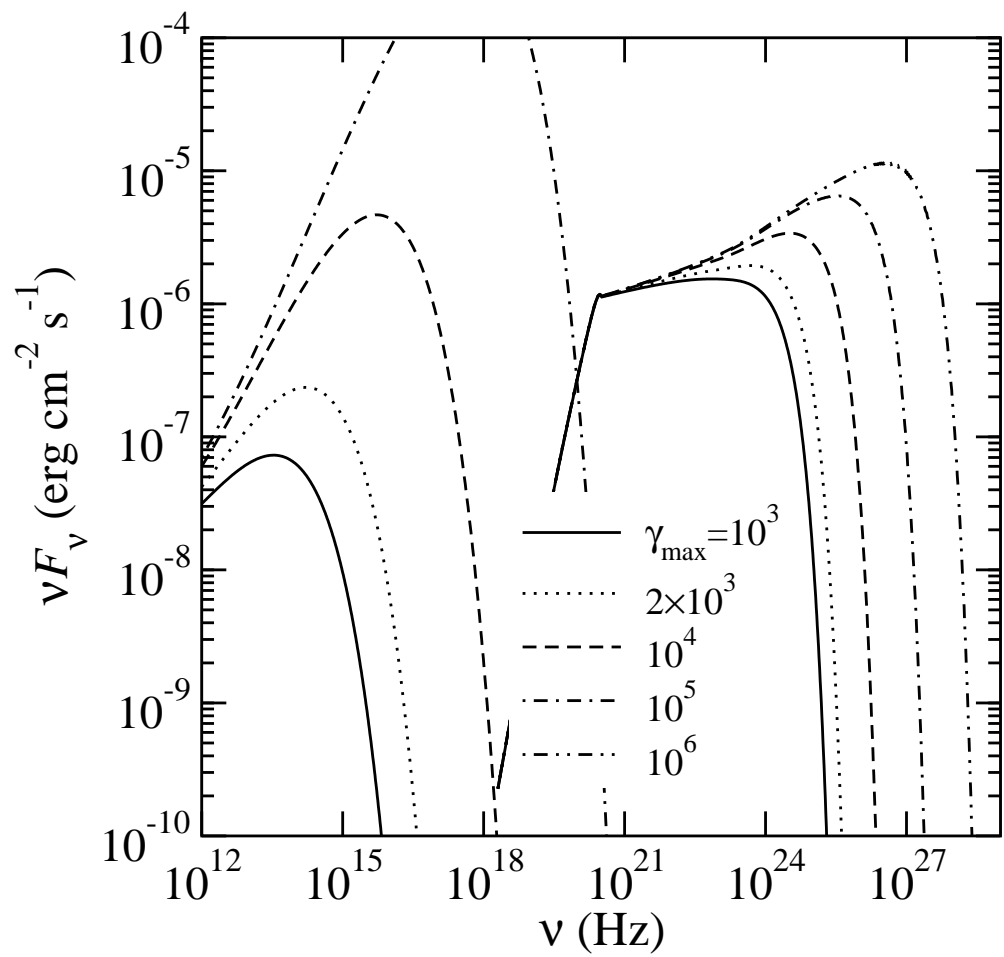}
\caption{Same as Fig. \ref{fig:ph-noesc-p2}, except $p = 1.8$.}
\label{fig:ph-noesc-p1.8}
\end{figure}

The observed photon spectra emitted by the electrons with $p =2$ shown
in Figure \ref{fig:el-noesc-p2-1.8} 
are presented in Figure \ref{fig:ph-noesc-p2} 
(the spectra from electrons with $p = 1.8$ given in Fig.
\ref{fig:el-noesc-p2-1.8} are shown in Fig. \ref{fig:ph-noesc-p1.8}).
It is assumed that
the emission region is in the moving blob with $\Gamma = 25$ at redshift
$z = 0.538$ (the redshift of 3C 279)
and that the size of the emission region is 
$R = 7 \times 10^{17} \, \text{cm}$ 
(the blob size appropriate for 3C 279 in our model; Kusunose et al. 2003).
The soft photon spectrum of the external photon source is monochromatic, 
with energy $x_0 m_e c^2 = 1.25 \, \text{keV}$.
Although the scattering cross section decreases
owing to the KN effects for $\gamma \gtrsim 0.1 x_0^{-1}$, 
there are still ample hard photons to be observed.
In these figures synchrotron spectra with $B = 0.3$ G are also shown.
The energy density of the magnetic field is
$3.6 \times 10^{-3}$ ergs cm$^{-3}$, which is to be compared with
the soft photon energy density of $8.3 \times 10^{-2}$ ergs cm$^{-3}$.
It is clear that when $\gamma_\mathrm{max}$ increases,
the synchrotron luminosity also increases.  
On the other hand, the Compton luminosity does not, because
of inefficient cooling in the KN regime.
Thus if the Compton cooling occurs mainly in the KN regime, 
the dominance of the synchrotron luminosity over 
the Compton luminosity does not mean that the magnetic energy 
density dominates other energy contents such as external soft photons.
It should be noted that 
the cooling rate of synchrotron emission is not taken into account
in the electron kinetic equation
and that the synchrotron spectra are shown only for the purpose of seeing 
the effect of different shapes of the electron spectra 
in the high-energy region.
For $\gamma_\mathrm{max} \gtrsim 10^3$, the synchrotron cooling
dominates the Compton cooling when the magnetic field is fixed at 0.3 G.
On the other hand, if $B \sim 0.02$ G, 
the synchrotron cooling time is one tenth of the Compton cooling time
for $\gamma \lesssim 10^5$.

As mentioned above, we solved equation (\ref{eq:kin}) at large values of
$\gamma$.   If we solve the equation at small values of $\gamma$ down to 1,
there should be infinite number of electrons at $\gamma = 1$,
and they should produce 
a peak at $h \nu \sim \Gamma^2 h \nu_0 \sim 31 \, \text{keV}$ 
in the photon spectrum, where $h \nu_0$ is the soft photon
energy in the rest frame of the soft photon source: 50 eV in our numerical
calculations.

From Figures \ref{fig:ph-noesc-p2} and \ref{fig:ph-noesc-p1.8}
it is found that the small differences in the electron spectrum
caused by the different values of $\gamma_\mathrm{max}$
result in different values of the power-law index of
the lower energy part of the synchrotron spectrum, or the part of
the spectrum below the synchrotron peak energy.
This might be used to determine observationally the KN effects on
the electron spectrum by the observations in radio through optical bands.

\begin{figure}
\epsscale{0.8}
\plotone{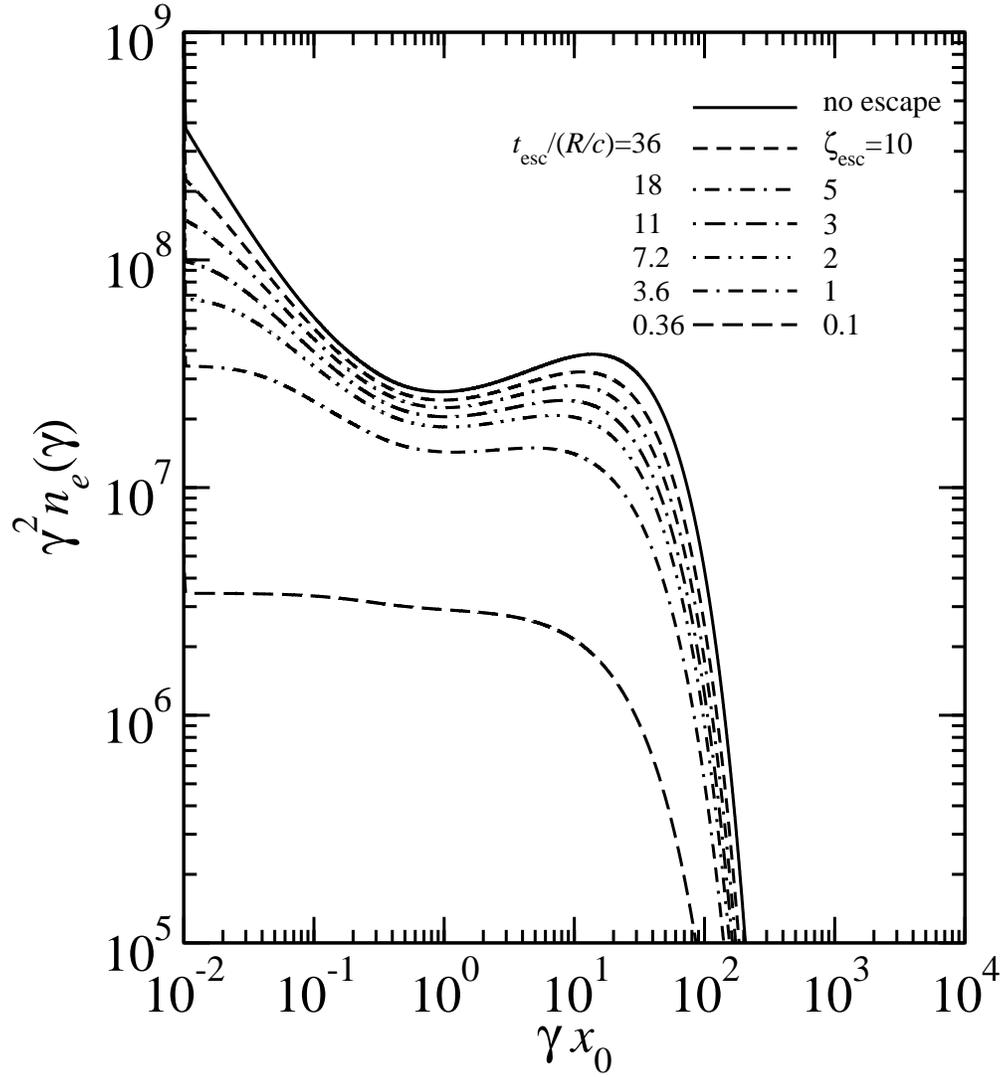}
\caption{Electron spectra with various values of escape time;
$\gamma_\mathrm{max} = 10^4$ and $p = 2$ are assumed for the injection
spectrum of electrons.
The soft photons are monochromatic with the energy 1.25 keV.
From upper to lower curves, $\zeta_\mathrm{esc} = \infty$ (no escape), 
10, 5, 3, 2, 1, and 0.1. 
}
\label{fig:el-esc-p2}
\end{figure}
\begin{figure}
\epsscale{0.8}
\plotone{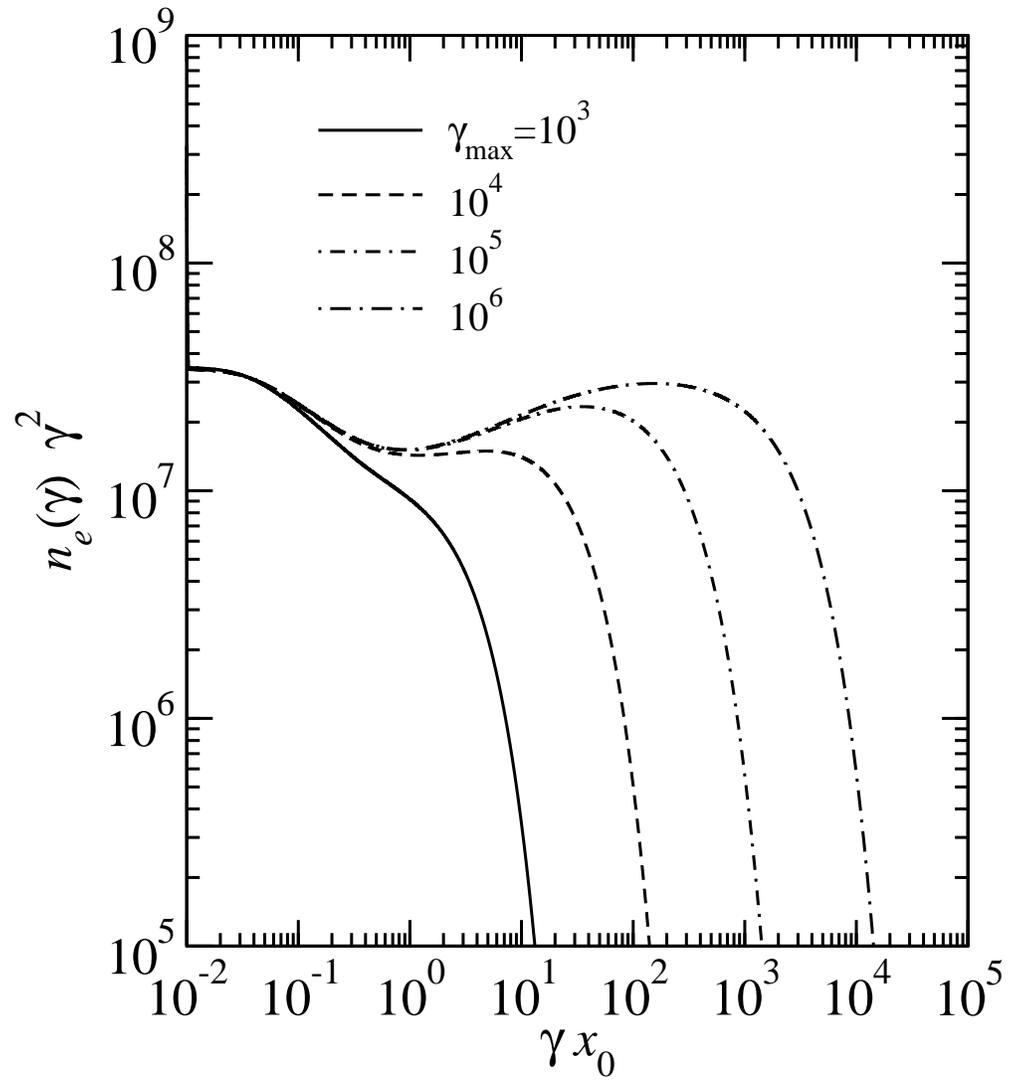}
\caption{Same as Fig. \ref{fig:el-esc-p2}, 
but for $\zeta_\mathrm{esc} = 1$ and 
the various values of $\gamma_\mathrm{max}$.}
\label{fig:el-esc-p2-3}
\end{figure}

The electron spectrum in the jet with electron escape from the emission region
is shown in 
Figure \ref{fig:el-esc-p2} for the different values of $\zeta_\mathrm{esc}$,
assuming that the soft photons are monochromatic.
In this figure, $\gamma_\mathrm{max} = 10^4$, $p = 2$,
and $m_e c^2 x_0 = 1.25 \, \text{keV}$ are assumed.
We assumed that $\zeta_\mathrm{esc} \propto t_\mathrm{esc}$, and
$\zeta_\mathrm{esc} = 1$ corresponds to $t_\mathrm{esc} = 3.6 R/c$
for the adopted parameter values.
When the escape time is shorter, the electron density decreases, 
the spectrum becomes flatter, and the peak energy of the bump decreases.
When $\zeta_\mathrm{esc} = 0.1$,
the shape of the electron spectrum is almost the same as that of the
injected electrons.
The electron spectra for the different values of $\gamma_\mathrm{max}$ are
shown in Figure \ref{fig:el-esc-p2-3} with $\zeta_\mathrm{esc} = 1$.
Even when the electron escape time is a few times the light crossing time
($R/c$), the energy density of high-energy electrons with 
$\gamma > x_0^{-1}$ is still large, 
and the emission from those electrons are expected to be as shown below.

\begin{figure}
\epsscale{0.8}
\plotone{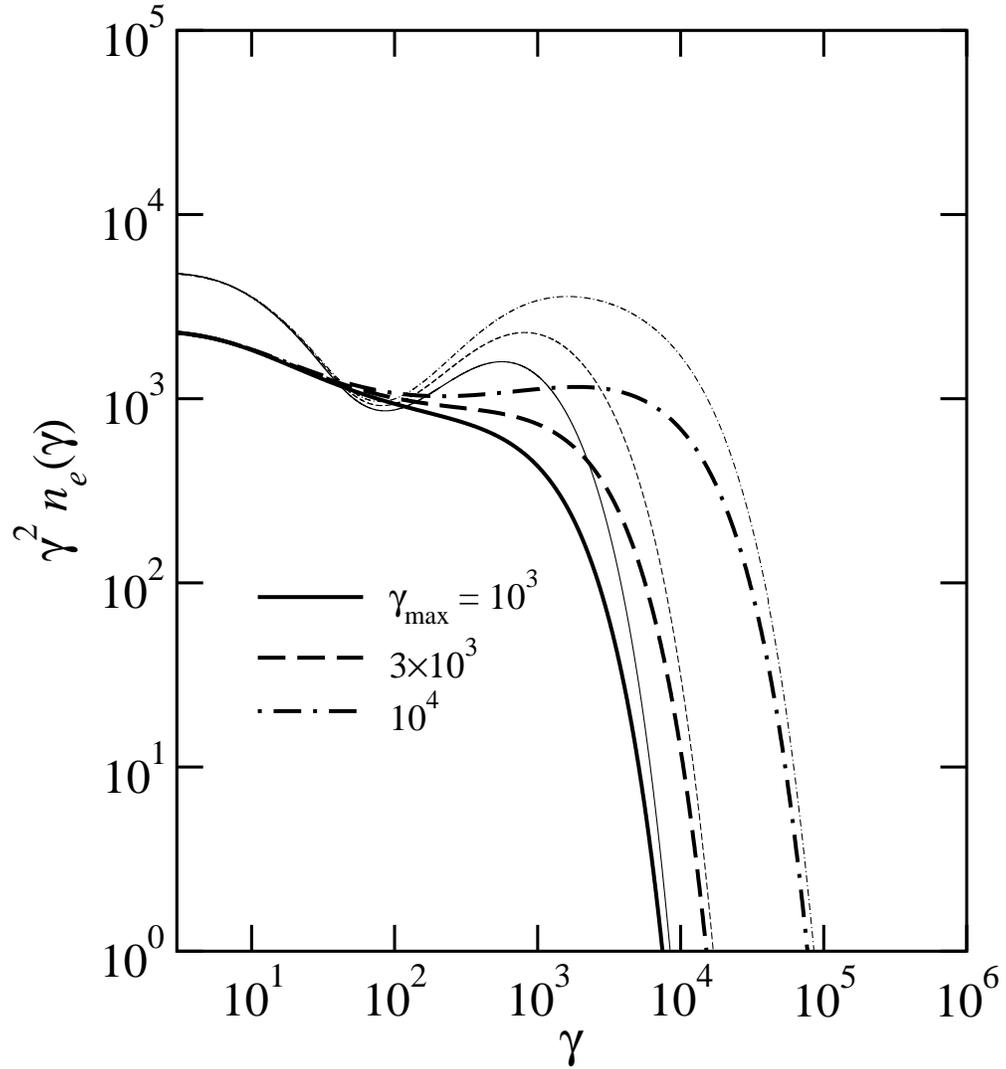}
\caption{Electron spectra with escape calculated with the KN
cross section ({\it thick lines}) 
are compared with those calculated by the Thomson approximation 
({\it thin lines}).
Electron escape time is $t_\mathrm{esc} = 1.5 \times 10^3$ s,
which corresponds to $\zeta_\mathrm{esc} = 1$ for the KN case.}
\label{fig:el-escape-bb-kn-th}
\end{figure}

The electron spectra with the Thomson approximation and those with
the KN cross section are compared 
in Figure \ref{fig:el-escape-bb-kn-th}.
The injected soft photons follow a blackbody distribution 
with a temperature of $kT = 1.25 \, \text{keV}$
and an energy density of $8.3 \times 10^{-2} \, \text{ergs} \, \text{cm}^{-3}$.
As with the case in which there is no electron escape,
the Thomson approximation results in more electrons in the high-energy region
than the calculations with the KN cross section.
In the Thomson approximation there is a dip at $\gamma \sim 100$,
which corresponds to $\gamma kT/(m_e c^2) \approx 0.24$.

\begin{figure}
\epsscale{0.8}
\plotone{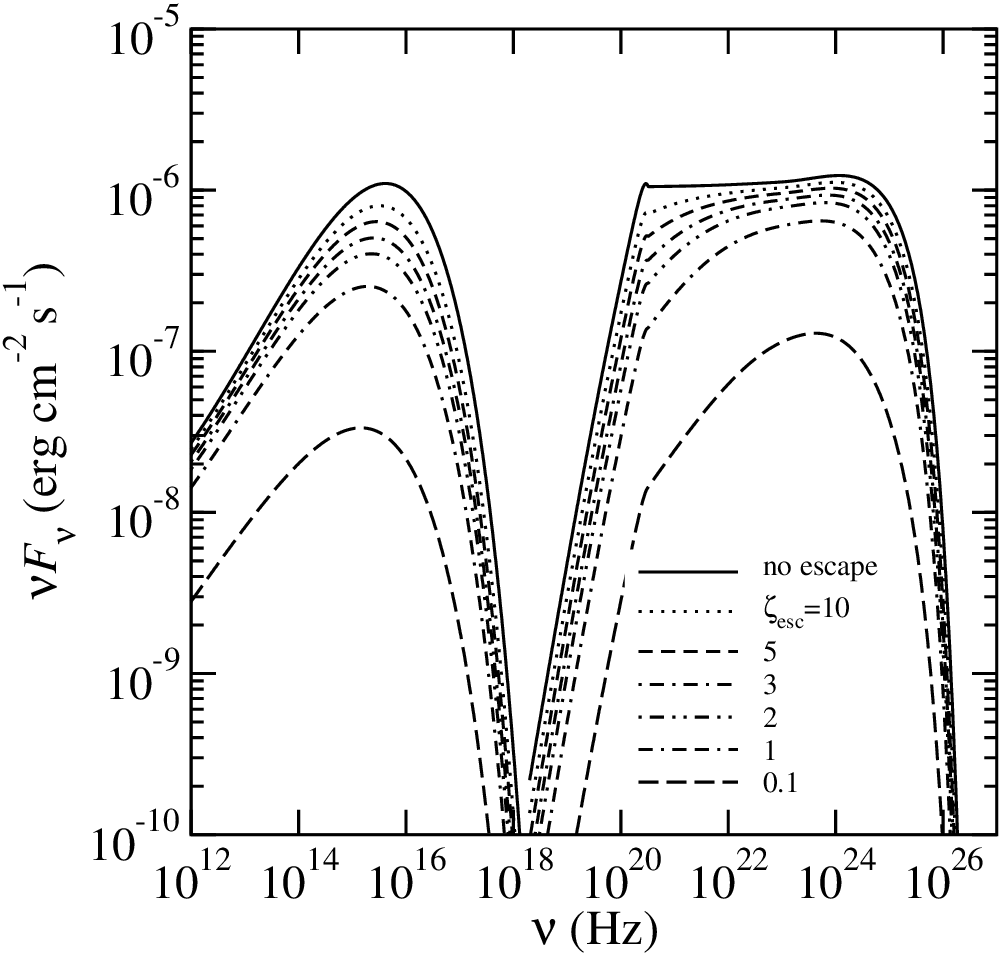}
\caption{Photon spectra emitted by the electrons shown 
in Fig. \ref{fig:el-esc-p2}.  Curves on the left are by synchrotron radiation
with $B = 0.3$ G, 
and curves on the right are by IC scattering.
}
\label{fig:ph-esc-p2-1}
\end{figure}
\begin{figure}
\epsscale{0.8}
\plotone{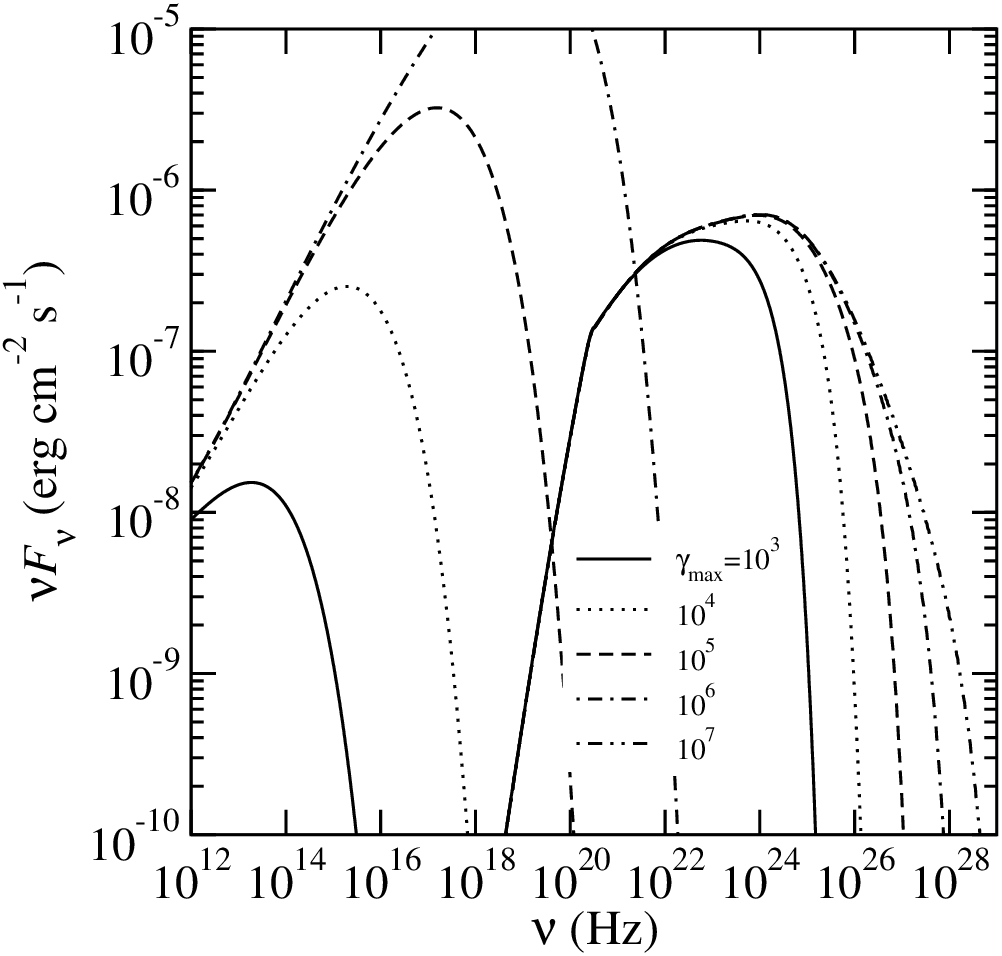}
\caption{Photon spectra emitted by the electrons shown 
in Fig. \ref{fig:el-esc-p2-3} ($\zeta_\mathrm{esc} =1$)
produced by IC scattering and synchrotron radiation with $B = 0.3$ G.
}
\label{fig:ph-esc-p2-3}
\end{figure}

The photon spectra from the jet with electron escape are shown in 
Figures \ref{fig:ph-esc-p2-1} and \ref{fig:ph-esc-p2-3}.
In both figures, the external soft photons are monochromatic.
Synchrotron emission with $B = 0.3$ G is also shown for comparison with
the IC emission. 
In Figure \ref{fig:ph-esc-p2-1}, $\gamma_\mathrm{max} = 10^4$,
$p = 2$, and various values of $\zeta_\mathrm{esc}$ are assumed.
When $\zeta_\mathrm{esc} = 0.1$, the Compton luminosity is higher than
the synchrotron luminosity.  
When $\zeta_\mathrm{esc}$ increases, both the Compton and synchrotron 
luminosities increase, while the spectral shape of both components 
changes as a result of the change in the electron energy 
distribution due to the KN effects.  
In Figure \ref{fig:ph-esc-p2-3}, $\zeta_\mathrm{esc} = 1$, $p = 2$,
and the different values of $\gamma_\mathrm{max}$ are assumed.
It is found that the peak of $\nu F_\nu$ of the Compton spectrum 
occurs at $h \nu \sim 10^3 m_e c^2$ for large enough values of 
$\gamma_\mathrm{max}$ ($\gtrsim 10^4$)
when $\zeta_\mathrm{esc} = 1$.
When $\zeta_\mathrm{esc}$ is larger, the peak of $\nu F_\nu$ occurs 
at a slightly larger value of photon energy.
The peak energy $h \nu \sim 10^3 m_e c^2$ is due to the electrons 
with $\gamma \sim 160$, i.e.,
$\gamma x_0 \sim 0.4$, which scatter seed photons in the Thomson regime
producing photons with energy $\sim \gamma^2 x_0 m_e c^2$.
Taking into account $z = 0.538$ and $\Gamma = 25$, the observed
peak energy appears at $h \nu \sim 10^3 m_e c^2$.
Because the synchrotron luminosity is larger than the Compton luminosity
for $\gamma_\mathrm{max} \gtrsim \text{a few} \,  \times \, 10^4$,
the properties of electron and photon spectra described here for
$\gamma_\mathrm{max} \gtrsim \text{a few} \,  \times \, 10^4$ 
are correct when $B$ is smaller than 0.3 G to keep the synchrotron
luminosity below the Compton luminosity.


\section{Summary}
\label{sec:sum}

We studied the effects of inverse Compton (IC) scattering 
in the Klein-Nishina (KN) regime, 
paying attention to the application to the gamma-ray emission 
from blazars.
For this purpose we solved the kinetic equation of electrons with electron 
escape, assuming that electron cooling is predominantly by 
IC scattering of external soft photons.
The photon spectrum was calculated, assuming that the emitting 
region is moving relativistically along the line of sight, i.e., 
emission from a relativistic jet is considered.
The seed soft photons of IC scattering are assumed to come from
external sources.
Because of the Doppler effect, the energy of the seed photons is higher for
electrons in the jet.  Then the KN effects appear for smaller
values of $\gamma$ compared with electrons in a stationary emitting region.
In our numerical calculations, the bulk Lorentz factor is set to
be 25, and the seed soft photon energy is 50 eV
in the frame of the soft photon source, which is based on our model
of 3C 279 \citep{kus03}.
For these parameters, electrons with $\gamma > 400$ scatter soft photons
in the KN regime.

When the escape of electrons can be neglected compared with the cooling,
the electron spectrum becomes harder than the injection spectrum
for $\gamma > x_0^{-1}$, as \cite{blu71} showed,
because of the less efficient cooling in the KN regime.
Although the cooling rate is smaller compared with that in the Thomson
regime, gamma rays with energy of $\sim \gamma_\mathrm{max}$ are amply emitted
by IC scattering.
When electron escape is taken into account, the number of high-energy
electrons decreases.  However, gamma rays in GeV region are still emitted
by IC scattering.

In astrophysical objects such as blazars, where IC scattering 
is a major radiative process, the spectrum of nonthermal electrons 
needs to be carefully considered.
Although a simple power-law spectrum of electrons might fit the observed
emission spectrum, it may not represent the physical conditions 
in the source.  The self-consistent treatment of photons and particles
should be taken into account. 
We finally remark on the effects of synchrotron radiation and
synchrotron-self-Compton scattering on the electron spectrum.
When the magnetic field is strong enough, these processes decrease
the number density of electrons with very large Lorentz factors,
and the upturn in the high-energy tail of the electron spectrum
becomes less prominent.
This effect was included in our previous paper \citep[][]{kus03},
where the KN effects are still apparent in the electron spectrum.


\acknowledgements

This work has been partially supported by Scientific Research Grants 
(M.K.:15037210; F.T.: 14079205 and 16540215) from 
the Ministry of Education, Culture, Sports, Science and Technology of Japan.



\begin{thebibliography}{}
\bibitem[Blumenthal \& Gould (1970)]{bg70}
Blumenthal, G. R., \& Gould, R. J. 1970, Rev. Mod. Phys., 42, 237

\bibitem[Blumenthal (1971)]{blu71}
Blumenthal, G. R. 1971, \prd, 3, 2308

\bibitem[Coppi \& Blandford (1990)]{cb90}
Coppi, P. S., \& Blandford, R. D. 1990, \mnras, 453, 469

\bibitem[Dermer \& Schlickeiser (1993)]{ds93}
Dermer, C. D., \& Schlickeiser, R. 1993, \apj, 416, 458

\bibitem[Ghisellini et al. (1996)]{gmd96}
Ghisellini, G., Maraschi, L., \& Dondi, L. 1996, \aaps, 120, 503,

\bibitem[Georganopoulos et al. (2001)]{gkm01}
Georganopoulos, M., Kirk, J. G., \& Mastichiadis, A. 2001, \apj, 561, 111
(erratum 604, 479 [2004])

\bibitem[Inoue \& Takahara (1996)]{it96} 
Inoue, S., \& Takahara, F. 1996, \apj, 463, 555 

\bibitem[Jones (1968)]{jones68} 
Jones, F. C. 1968, Phys. Rev., 167, 1159

\bibitem[Kusunose et al. (2003)]{kus03}
Kusunose, M., Takahara, F., \& Kato, T. 2003, \apjl, 592, L5

\bibitem[Lightman \& Zdziarski (1987)]{lz87} 
Lightman, A. P., \& Zdziarski, A. A. 1987, \apj, 319, 643

\bibitem[M\"{u}cke et al. (2003)]{muc03}
M\"{u}cke, A., Protheroe, R. J., Engel, R., Rachen, J. P., \&
Stanev, T. 2003, Astropart. Phys., 18, 593

\bibitem[Rybicki \& Lightman (1979)]{rl79}
Rybicki, G. B., \& Lightman, A. P. 1979, 
Radiative Processes in Astrophysics, New York: Wiley

\bibitem[Sikora et al. (1994)]{sbr94}
Sikora, M., Begelman, M. C., \& Rees, M. J. 1994, \apj, 421, 153 

\bibitem[Zdziarski (1988)]{zdz88}
Zdziarski, A. A. 1988, \apj, 335, 786

\bibitem[Zdziarski (1989)]{zdz89}
------------. 1989, \apj, 342, 1108

\bibitem[Zdziarski \& Krolik (1993)]{zk93}
Zdziarski, A. A., \& Krolik, J. H., \apjl, 409, L33
\end{thebibliography}
\end{document}